\documentstyle[aps,multicol,epsfig,psfig,pre,colordvi]{revtex}  
\def\beq{\begin{equation}}
\def\eeq{\end{equation}}
\def\beqa{\begin{eqnarray}}
\def\eeqa{\end{eqnarray}}

\def\bdi{\begin{displaymath}}
\def\edi{\end{displaymath}}

\title{Stretching of a polymer below the $\Theta$ point}
\author{D. Marenduzzo$^{1}$, A. Maritan$^{1,2}$, A. Rosa$^{1}$, F. Seno$^{3}$} 
\address{$^1$ International School for Advanced Studies (SISSA) 
  and INFM, 
  Via Beirut 2-4, 34014 Trieste, Italy \\ 
  $^2$ The Abdus Salam International Center for Theoretical Physics 
  (ICTP),  
  Strada Costiera 11, 34100 Trieste, Italy \\ 
  $^3$ INFM and Dipartimento  di Fisica - Universit\`a di 
  Padova, Via Marzolo 8, 35131 Padova , Italy\\ } 
\begin{document}

\tightenlines

\maketitle

\begin{abstract}
The unfolding of a polymer below the $\theta$ point when pulled by an
external force is studied both in $d=2$ on the lattice and in $d=3$
off lattice. A ground state analysis of finite length chains shows that
the globule unfolds via multiple steps, corresponding to transitions
between different minima, in both cases. In the infinite length limit,
these intermediate minima have a qualitative effect only in $d=2$. The
phase diagram in $d=2$ is determined using transfer matrix techniques. 
Energy-entropy and renormalization group arguments are given which predict a
qualitatively correct phase diagram and a change of the order of the
transition from $d=2$ to $d=3$. 
\end{abstract}


\begin{multicols}{2} 

The recent refinements in experimental techniques employing optical 
tweezers\cite{optical}, 
atomic force microscopes\cite{afm}, soft microneedles\cite{needles}, 
make potentially  
possible for researchers to monitor the behaviour under tension and stress
of various biopolymers and then to elucidate the mechanism of 
some force-driven phase transitions occurring at the single molecule 
level, such as the unfolding of the giant titine
protein\cite{titin}, the stretching of single collapsed 
DNA molecules\cite{bustamante,plateaus}, the unzipping of RNA and DNA\cite{optical}. 
Theoretically, on the other hand, quite a few statistical
mechanics models have been subsequently proposed in order to
explain these experimental results and to identify the physical
mechanisms behind these phase 
transitions\cite{halperin,thir,vilgis,shak,grassberger,jpa,hoang,theory,rientranza}. 
In this work we focus on two theoretical models aimed at 
understanding the unfolding behaviour of polymers in a 
poor\cite{vanderzande} solvent (globules), i.e. below the 
theta temperature, $T_{\theta}$.

This is motivated by the vast 
number of experimental results in the literature. 
In particular, various force vs elongation ($f$ vs $x$) curves have been 
recorded in experiments studying these phenomena\cite{bustamante,plateaus}. 
In most cases a curve consisting of three distinct regimes, 
and in particular displaying a plateau for intermediate 
stretch\cite{bustamante,plateaus}, have been observed; 
whereas in a few examples, for shorter polymers, a stick-release pattern with 
histeresis has been found. The first observation is in good agreement with the mean 
field theory proposed in Ref.\cite{halperin}, and the plateau 
strongly suggests the presence of a first order phase transition
(see also the recent measurements in \cite{plateaus}). 
Our results suggest that there might be more than one 
possible shape for the $f$ vs. $x$ curves according to $d$, the spatial 
dimension and to polymer length, so that a
different behaviour occurs when mean field theory\cite{note1} is qualitatively incorrect.
These models have a remarkable interest even on a purely theoretical ground. 
First, the numerical study recently performed in \cite{grassberger} 
has suggested the possibility that the transition 
is second order in $d=2$ and first order in $d=3$. 
This has been confirmed to some extent in a study of the 
model on a hierarchical lattice with fractal dimension
two\cite{jpa}. The $d=2$ case is important as 
it is below the upper critical dimension for theta collapse and
mean field predictions may well be incorrect. A thourough analysis
and a clear physical mechanism underlying the difference of the
nature of the transition as $d$ changes are then needed. 
Second, the mean field analysis of Ref.\cite{shak} has 
suggested there could be a re-entrant region in the phase diagram 
for low temperature similar to what happens for DNA unzipping\cite{rientranza}. 
However the exact results in\cite{jpa} prove mean field is not valid in $d=2$. 

Then, our aim here is to describe theoretically the
unfolding transition of globules not relying on the mean field 
approximation. We first characterize the evolution of the ground states of a
finite polymer as the pulling force increases. This will help 
to understand the thermodynamics. In particular, we compute the phase diagram in the 
temperature--force $(T,f)$-plane in $d=2$ on the lattice, where we can use the transfer 
matrix method (Fig. 2) together with exact enumerations (Fig. 1).


We begin by considering a self-avoiding walk (SAW) on the square 
lattice with {\it fixed} origin. The model partition function (generating
function) in the canonical ensemble in which $T$ and $f$, the stretching force, are 
fixed is: 
\begin{equation}\label{tmpartfunc}
Z_N(f,T) = \sum_{\mathcal {C}}
e^{-\beta H(\mathcal{C})} = \sum_{b,\mathbf{R}} w_{o\mathbf{R}}
(N, b) e^{\beta(b\epsilon+fR_x)}
\end{equation}
where $N$ is the number of monomers (including the origin) of the SAW, $H(\mathcal{C})$ 
(referred to a configuration $\mathcal{C}$, i.e. to a SAW) is the energy of a SAW,
$b$ is the number of pairs of neighboring occupied sites not adiacent along the chain, 
$\beta\equiv\frac{1}{T}$ is the inverse temperature, $w_{o\mathbf{R}}(N, b)$ 
is the number of configurations of a SAW with fixed origin $o$ and end-to-end distance 
$\mathbf{R}$, of lenght $N$ and  $b$ contacts, $R_x$ is the projection along the force 
direction ($x$ axis) of $\mathbf{R}$. We take both the Boltzmann constant and $\epsilon$
equal to one.

When $T$ is low, one may look for the ground states among the rectangles of sides $L_x$ 
and $L_y$ that are completely covered by the SAW (in other words such that 
$L_x L_y=N$, where we neglect the small effects arising when this 
rectangle cannot be constructed with both $L_{x,y}$ integers).
The energy of this rectangular hamiltonian walk with a non zero $f$ 
is $-H(N\equiv L_xL_y,L_x)=N-L_x-\frac{N}{L_x}+1+f(L_x-1)$.
The minimum of $H(N,L_x)$ for given $N$ with respect to $L_x$ yields the most stable
configuration for various values of $f$. The minimum occurs for an
$f$-dependent value of $L_x^0\equiv L_x^0(N,f)$, namely $L_x^0(N,f)=
\sqrt{\frac{N}{1-f}}$. For any $f<1$ one has a compact configuration. 
However, when the critical value $f=1$ (for $T=0$) is reached all integer values of $L_x$
from $N$ (stretched coil) to $L_x\sim N^{1/2}$ (compact globule) become degenerate for 
large $N$. Note that this does not hold in $d=3$ where it is well known that there is
a Rayleigh instability in the thermodynamic limit\cite{plateaus,halperin,grassberger}.
This can be seen by comparing the globule energy - which in $d=3$ is $2N$ 
in the large $N$ limit - with the energy of a parallelepiped, 
with elongation along $\vec f$ equal to $L_x$ and with edges $L_y=L_z$ in the
perpendicular plane. The force above which (for $T=0$) the parallepiped is a better 
ground state than the three-dimensional globule is $2L_y$ ($L_y \ll N^{1/3}$), 
and there is no longer any degeneracy at the critical force $f_c=2$ (at $T=0$).

\begin{figure}
\centerline{\psfig{figure=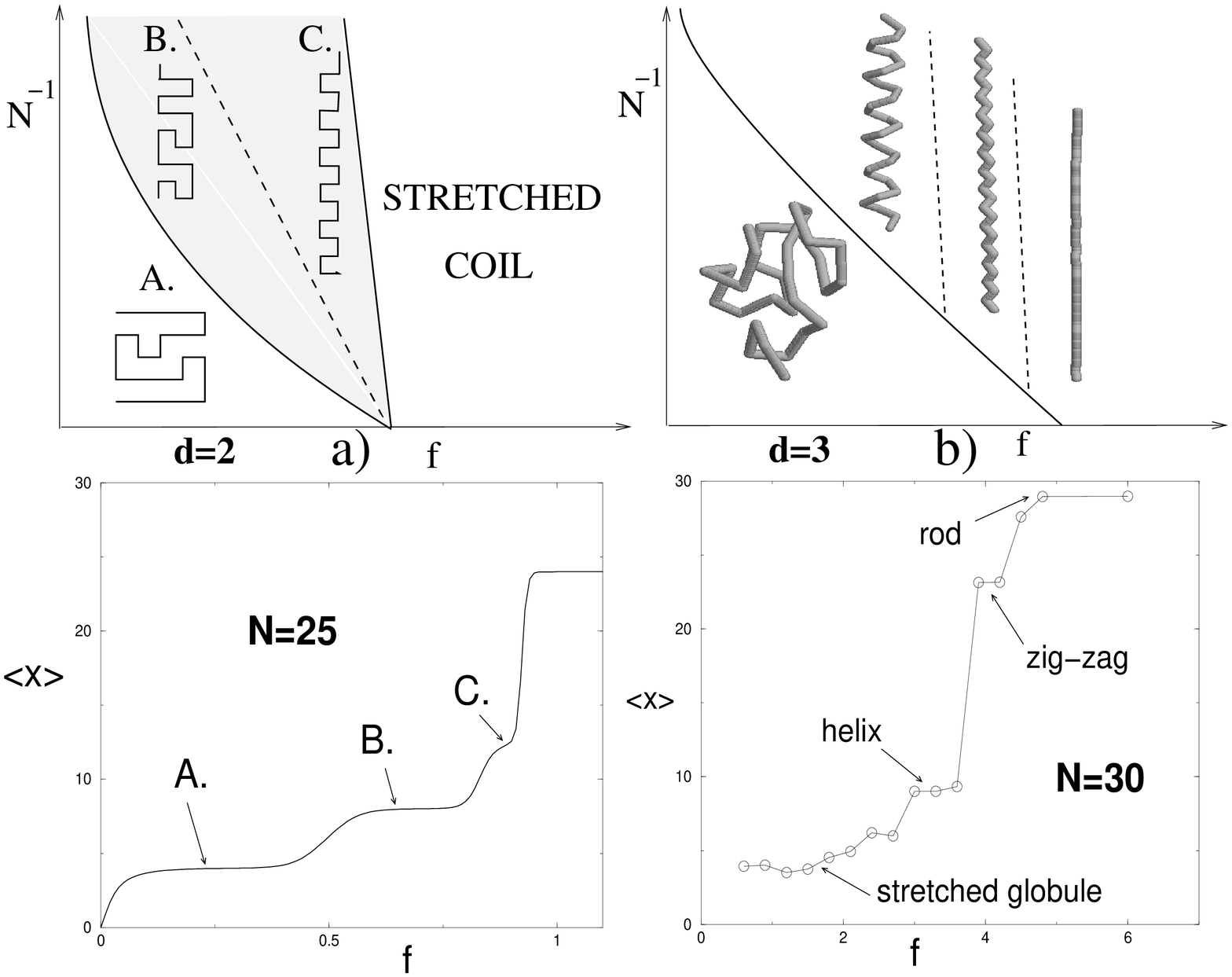,width=3.in}}
\caption{a) $d=2$ Schematic diagram of minima for polymers of different sizes 
(top) and low $T$ $<x>$ vs. $f$ curve for $N=25$ (bottom) found with 
exact enumerations in $d=2$. b) d=3: Same as in a), except that the
$<x>$ vs. $f$ curve (bottom) is for $N=30$ and is found by simulated annealing.}
\label{stretch}
\end{figure}

In Fig. \ref{stretch}a we sketch the situation in $d=2$.
The minima hierarchy, shown in the shaded area in the top panel, affects 
the low $T$ region of the $<x>$ (average elongation) vs. $f$ curves for 
finite length 
(bottom panel). However, only one transition survives in the 
large $N$ limit and represents a true phase transition (as represented in Fig. 
\ref{stretch}a
by the shaded wedge ending in just one point in the $N^{-1}=0$ axis). Similarly,
when $T$ is raised the multi-step character of the $<x>$ vs. $f$ curves is lost 
due to fluctuations which blur the ground state dominance in the
partition.  In Fig. \ref{stretch}b we show the analogous picture for a $3d-$model 
 discussed below. 
Short chains in $d=2$ and $3$ behave similarly whereas in the infinite length limit
the $d=3$-case shows an abrupt unfolding transition.


We now turn to the thermodynamic behaviour of the SAW model on a
square lattice.  We use the transfer matrix (TM) technique,
following Refs.\cite{Derrida,Antonio e Flavio,Saleur}.

Let us introduce briefly the principal features of the TM approach: the
partition function of a polymer of $N$ sites is given by Eq. (\ref{tmpartfunc}), 
with $f=0$. In the thermodynamic limit ($N \rightarrow \infty$) we expect that 
$Z_N(T) \sim [\mu(T)]^N$, then the free energy per monomer ${\mathcal F}$ is simply,
 ${\mathcal F} = -T\log\mu(T)$. It is more convenient\cite{Derrida} to introduce the 
following {\it generating function} ($z$ is the step fugacity)
\begin{equation}\label{tmgenfunc}
g_{o\mathbf{R}} = \sum_{N, b}z^N e^{\beta b}w_{o\mathbf{R}}(N, b)
\end{equation}
It is known that for $z < z^c(T) = 1/\mu(T)$, the
inverse SAW connectivity, $g_{o\mathbf{R}} \sim
\exp(-R_x/\xi(z, T))$, where $\xi(z, T)$ is the
{\it correlation length} and $R_x$ is the projection of ${\mathbf R}$ along $x$. 
We study the stretching of an interacting SAW in a strip of
finite size $L$ along $y$ and infinite length along $x$. 
It is possible to define\cite{Derrida} an $L$-dependent correlation
length $\xi_L(z, T)$ {\it via} the formula
$\xi_L(z, T) = -\frac{1}{\log\lambda_L(z, T)}$, 
where $\lambda_L(z, T)$ is the {\it largest} eigenvalue of the transfer
matrix, that equals $1$ at $z = z^c_L(T)$. We apply the {\it
phenomenological  renormalization}, to find successive estimates for 
 $z_c(T) = \lim_{L \rightarrow \infty} z_L^c(T)$.
Including the force via Eq. (\ref{tmpartfunc}), the equation for the critical 
force $f_c(T)$ is then ideally found via:
\begin{equation}\label{critforce}
f_c=-T\lim_{z\to z_c(T)^{-}}\lim_{L\to\infty} \log{\lambda_L}  .
\end{equation}
The order of the limits in Eq. (\ref{critforce}) and a correct choice 
of the boundary conditions (see below) are crucial \cite{note}.

In Fig. \ref{phasediag} the {\it phase diagram} for the stretched 
interacting SAW is shown. With the TM, a right choice of the {\it boundary conditions} 
is needed\cite{Derrida,Antonio e Flavio,Saleur}.   
We have used both {\it periodic} (PBC) and {\it free boundary conditions}
(FBC). PBC have been employed to get the best
estimate of $z_c(T)$ through phenomenological  renormalization. This
value is then used  with FBC, to find the correct $L$-dependent critical 
eigenvalue  $\lambda_L(z_c(T), T)$.  Finally,
adopting the extrapolation algorithm of \cite{Henkel and Schutz},
$\lim_{L \rightarrow \infty} \lambda_L(z_c(T), T)$  is obtained 
which, through Eq. (\ref{critforce}), allows to get the  phase diagram.
As expected, in the case of FBC, there are oscillations in data going from 
odd $L$ to even $L$. As usual in this context, a separated analysis of
even and odd $L$ data was necessary for obtaining a better convergence
(see Fig. \ref{phasediag}). One point on the transition line obtained
previously in \cite{grassberger}, is recovered here.
 
One can get an approximate description of the transition if
one requires that the globule and coil phases coexist. The globule free
energy is easily estimated in terms of hamiltonian walks\cite{vanderzande}. 
On a square lattice the energy is simply given by
minus the length of the polymer whereas the entropy is given in terms of the
number of hamiltonian walks which grows exponentially with $N$\cite{vanderzande}. 
Thus the globule free energy per monomer is ${\mathcal F}_g=-1-T\log(4/e)$ where we have
used the accurate mean field estimate of the entropy as given in \cite{lise}.
The coil free energy ${\mathcal F}_c$ is approximated as that of an unconstrained
random walk in presence of a pulling force and contacts are neclected.
We thus get ${\mathcal F}_c=-\log(2(1+\cosh(\beta f)))$. At coexistence one finds 
$f_c(T) = T{\cosh}^{-1}(2\exp(1/T-1)-1)$ 
(the continuos curve in Fig. \ref{phasediag}).

We note that $f_c(0)=1$ is the exact result and at low $T$ the phase
diagram displays a reentrant region. As $T\to T_{\theta}$, $f$
approaches $0$ rather smoothly, consistently with the prediction
$f\sim(T_{\theta}-T)^ {\frac{\nu_{\theta}}{\phi_{\theta}}=\frac{4}{3}}$
\cite{jpa}. Within the  TM approach one can also infer the order of
the transition. To do this we observe that, 
if for $z\to z_c(T)^{-}$:
\beq\label{order} 
\lambda(z_c(T),T) - \lambda(z,T) \sim (z_c(T)-z)^{\Delta},
\eeq
then $\Delta<1$ ($\Delta=1$) means a second (first) order transition,
as $<x>\sim(f-f_c)^{\frac{1}{\Delta}-1}$ for $f\sim f_c$.
Our data at not too low $T$ are compatible with a second order 
transition (inset of Fig. \ref{phasediag}).

Inspired by exact renormalization group (RG) on the
Sierpinski lattices \cite{jpa}, and on approximate RG in $d-$ dimensional
lattices \cite{physicaa}, we propose the
following simplified real space RG which rationalizes our results. The RG
recursions relations can be written for the generating function
representing polymers traversing  a hypercube of linear size $1$ once,
$A$, and $2^{d-1}$-times, $B$ (Fig. 3). The terms $A$ and $B$ represent parts of the
chain which are in the coil and globular state respectively. The
recursion relations can be calculated as in
Ref.\cite{dhar} or by enumerating the SAWs on $2\times2$ or $2\times2\times2$ 
cells as in \cite{physicaa}.  To leading order in $A$ at $0\ne T\ll T_{\theta}$ 
\beq\label{rsrg} 
A'=A^2+\alpha(d)A^2B^2, \quad B'=\gamma(d)B^{2^d}, 
\eeq 
where $\alpha(d)$ and $\gamma(d)$ are $d-$dependent constants. 
There are three fixed points in the flux in Eq. \ref{rsrg}:
$A=0$,  $B=B^*=\gamma(d)^{-\frac{1}{2^d-1}}$ corresponds to the
globular phase,  $(A,B)=(1,0)$ to the coil phase, while the last
non-trivial fixed point $A=A^*=\frac{1}{1+\alpha(d)(B^*)^2}$,
$B=B^*$  characterizes the unfolding transition.  The value of the term
$\alpha(d)$ affects the behaviour of the RSRG flux near the fixed point
$(A^*,B^*)$. One can see\cite{note2} that $\alpha(d)\ne 0$ if $d=2$ and 
is $0$ in $d>2$. When $\alpha(d)\ne 0$ (i.e. in $d=2$), the RG flux is smooth near
$(A^*,B^*)$ and the critical fugacity near $f=f_c(T)$ behaves as 
$z_c(f_c(T),T)-z_c(f,T)\sim (f-f_c(T))^{2}$ signalling a second order
transition  with $\Delta=1/2$ in Eq.\ref{order}. On the other hand 
when $\alpha(d)=0$ (i.e. $d>2$), the transition is first order and
two-state like.  The presence of the mixed term in Eq. (\ref{rsrg}) is
crucial and enhances the entropy  of the coil phase since it
contributes to $A'$. Consequently, the $d=2$ 
two-state approximation in Fig. \ref{phasediag} gives a transition
line which is higher than the
numerical result for $0\ne T\ll T_{\theta}$. 
The entropy gain in the stretched coil, as $T\to0$, is
hampered as it costs a finite surface energy (dominant as $T\to 0$)
to change locally an elongated  globular region  into a coil and
vice-versa. This is why the solid curve in Fig. 2 matches our
numerics as $T\to 0$.

Let us now discuss the $d=3$ case. The model we used is
the freely jointed chain (FJC) (see e.g. \cite{vanderzande}) 
in the  continuum (off lattice). The FJC is subject to a
compacting pairwise attractive  potential between non-consecutive beads
and to a stretching force $\vec f$ at the extrema.  The pairwise
potential is chosen to be an asymmetric square well with a hard core
radius, $2R_{hc}$, which acts between non-consecutive  beads along the
chain, and an attraction range $R_1 >2R_{hc}$, i.e. the distance up to
which  the particles interact. We have checked that the results
reported in the following do not appreciably depend on the
two-body potential details. The parameters we have used to generate the
configurations shown in Fig. \ref{stretch}b (top panel)  are
$R_{hc}=0.6$, $R_1=1.6$ where the unit length is the distance
between sucessive beads along the chain.

The ground states of  short chains (up to $N=30$ bead long) has been
determined by performing simulated annealing employing Monte-Carlo
dynamical simulations. The FJC is evolved dynamically by means of three
sets of moves: the pivot, reptation and crankshaft moves\cite{sokal}.
We lower $T$ during the simulation according to a 
standard annealing schedule. We found evidence also in this $d=3$
case that the unfolding of a finite length homopolymer proceeds 
in a multistep fashion. The collapsed globule
first orients itself along the  pulling force as soon as there is a
nonzero $f$. At larger $f$  the globule is slightly elongated
(much less than in the $d=2$ case) and after this a helix forms
followed by  a zig-zag curve and finally by a stretched coil.  This
succession of minima, shown in fig. \ref{stretch}b is intriguing for a 
two-fold reason: firstly because it suggests that helices, one of the
well known building block of proteins, come out rather naturally as
one of a few stable minima of a homopolymer in a poor solvent
subject to a finite stretching force; secondly because the unfolding
transition of a finite polymer in this model appears to be markedly
different from a globule-to-coil (two-state) transition.  
The mean field picture of an all or none transition is recovered for infinite length. 
The situation is depicted in Fig. \ref{stretch}b. 
Since in $d=3$ the transition is first order, the mean field treatment is correct in the
thermodynamic limit. The fact that helices become better ground
states than compact globules could be  easily verifiable in experiments
done in the fixed force ensemble.  These can now be confidently
performed, with not more substantial difficulties than  the more
conventional ones, performed in the fixed stretch ensemble. Indeed the
stick-release pattern reported in Ref.\cite{bustamante} shows several 
different peaks in the $f$ vs. $<x>$ curve, and is compatible with a
`multi-step  transition' in which the globule undergoes more than one
conformational changes  (signalled by the peaks) during the unfolding.
Intriguingly, this pattern is  reported for smaller polymers, a fact
which would be in agreements with our findings; whereas for large $N$
this effect is much less important and the curves have a single (dominant)
plateau.

Helices appear as ground states for a potential consisting of a force term and a  
hydrophobic contact potential because they are both elongated and offer a good
shielding from the outside solvent to the monomers lying in its
interior. Note also that helices arise as optimal states of  tubes of
non-zero thickness subject to compaction\cite{tube}. If a worm-like
chain  is used instead of the FJC, the picture should not change, because the elongated
states such as the helices and the zig-zag are stiffer than the
collapsed globule (sharp corners between successive monomers are necessary in this 
state) and so should be even more favoured as $f$ is increased.

In conclusion, we have discussed the unfolding transition of a
homopolymer under the action of an external pulling force in $d=2$ on
the lattice and in $d=3$ off lattice. A ground  state analysis shows that
for finite length polymers, the unfolding is not abrupt, rather it occurs
via a multi step sequence of states. These are more elongated than the
globule and  make more contacts than a coil. In $d=3$ 
helices arise naturally as  ground states at intermediate forces. For
infinite polymers, on the other hand, the situation is different: in
$d=3$ the intermediate ground states disappear due to the Rayleigh
instability, and the transition is effectively two-state, whereas in
$d=2$ they survive in the thermodynamic limit. Indeed, from the transfer
matrix results, it is apparent that the mean field hypothesis is 
incorrect in $d=2$ even at rather low $T$: the transition is
second order as also found in Ref.  \cite{grassberger} although the
prediction of the re-entrant region agrees with the TM results.
Furthermore a renormalization group based  argument is in agreement
with this picture. 

This work was supported by Cofin 2001 and FISR.

\begin{figure}
\centerline{\psfig{figure=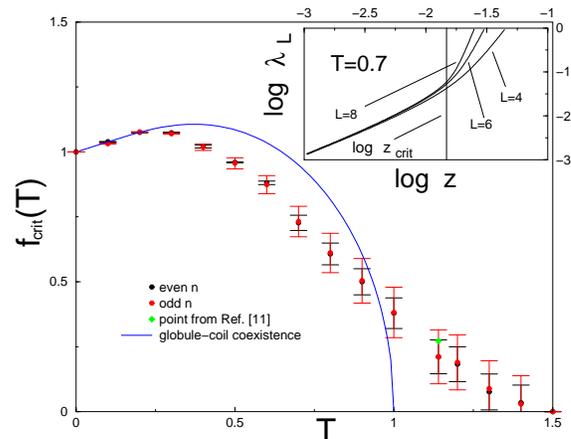,angle=270,width=3.in}}
\narrowtext
\caption{Phase diagram for the stretching of a SAW on a $2d$-lattice, 
obtained with the TM technique. Inset: plot of $\log{\lambda}$ vs. 
$\log{z}$ for $T=0.7$.}
\label{phasediag}
\end{figure}

\begin{figure}
\centerline{\psfig{figure=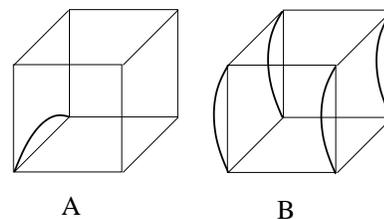,width=2.in}}
\narrowtext
\caption{Representation of the generating functions $A$ and $B$ in the text
(for $d=3$) at a generic level of iteration in the RG. 
Thick lines represent portions of the SAW that enter through a vertex
and go out from another one.}
\label{rsrgfigure}
\end{figure}

\end{multicols}


\begin{thebibliography}{99}
\bibitem{note1} Note that we use `mean field theory' in a broader
sense, including approaches which consider the relative stability of the globule and
the coil only (see also \cite{shak}). Mean field theories should be all 
equivalent as regards universality.
\bibitem{note} This and other subtleties of the TM approach
(unnecessary here) will be addressed in detail
elsewhere.
\bibitem{note2} These facts can be proved exactly in a family of Sierpinski gaskets
(see \cite{jpa} for an example), where $d$ is the fractal dimension. On conventional
lattices, an approximate RG treatment as in \cite{physicaa} gives qualitatively
the same behaviour. Details will be given in a future work.
\end{thebibliography}
\end{document}